\documentstyle[12pt]{article}
\begin{document}
\title{The Emergence of the Planck Scale}
\author{B.G. Sidharth$^*$\\
Centre for Applicable Mathematics \& Computer Sciences\\
B.M. Birla Science Centre, Adarsh Nagar, Hyderabad - 500 063 (India)}
\date{}
\maketitle
\footnotetext{$^*$Email:birlasc@hd1.vsnl.net.in; birlard@ap.nic.in}
\begin{abstract}
In this paper we first observe some interesting parallels between Planck
scale considerations and elementary particle Compton wavelength scale
considerations, particularly in the context of Wheeler's space time foam and
a space time arising out of a stochastic random heap of elementary particles
discussed in previous papers. These parallels lead to a semi qualitative
picture which shows how the short lived Planck scale arises from the
Compton wavelength considerations. Finally all this is quantified.
\end{abstract}
\section{Introduction}
About a century ago Max Planck had pointed out that the quantity
$\left(\frac{\hbar G}{c^3}\right)^{\frac{1}{2}} \sim 10^{-33}cms$ is a fundamental length. This so called
Planck length ties up Quantum Mechanics, Gravitation and Special Relativity
and leads to the Planck mass $\sim 10^{-5}gms$. It is but natural that the
Planck length has played a crucial role in Quantum Gravity as also in
String Theory which includes a description of Gravitation, unlike Quantum
Theory or Quantum Field Theory.\\
It turns out to be the scale at which we have no longer the smooth space
time of Classical Theory and Quantum Theory, but rather we have the space
time foam of Wheeler\cite{r1,r2}. This is inextricably linked with gravitational
collapse which has been described by Wheeler as "The greatest crisis of
Physics". As he puts it, "These are small scale fluctuations telling one that
something like gravitational collapse is taking place everywhere in space
and all the time; that gravitational collapse is in effect perpetually being
done and undone .... at the Planck scale of distances." In this space time
foam, worm holes and non local effects abound.\\
On the other hand there is also a stochastic fluctuational picture of space time that deals with
phenomena at the Compton wavelength scale and leads to meaningful physics
and cosmology including a unified description of gravitation and
electromagnetism consistent with observation\cite{r3,r4,r5,r6,r7}. In this picture, space time has been considered to be
a random heap\cite{r8} of elementary particles. If
we consider a typical elementary particle to be a pion with Compton
wavelength $l$, then the above picture leads to a dispersion length in the
Gaussian distribution $\sim \sqrt{N}l, N \sim 10^{80}$ being the number
of elementary particles in the universe, this being the correct
dimension of the universe itself.\\
We will now show a parallel between the Planck length considerations and
the Compton wavelength considerations referred to above, which will then
show us how the Planck length considerations emerge.
\section{The Emergence of the Planck Scale}
We first show the parallels between the Compton wavelength picture and the Planck
length picture. We note that in the former scenario, particles are
fluctuationally created at the Compton wavelength from a background pre
space time Zero Point Field (ZPF) of the kind considered in stochastic
electrodynamics\cite{r9,r10}. The energy content in terms of the magnetic
field of such a particle is given by (Cf.ref.\cite{r3})
\begin{equation}
\Delta B \sim \frac{(\hbar c)^{1/2}}{L^2}\label{e1}
\end{equation}
where $L$ is the dimension under consideration, which in this case is of
the order of the particle's Compton wavelength. We note that in (\ref{e1})
if $\hbar c$ or equivalently $137 e^2$ is replaced by its gravitational
counterpart, namely $137 Gm^2$ then we get, as in the fluctuation of the
metric\cite{r1},
\begin{equation}
\Delta g \sim \frac{L_P}{L}\label{e2}
\end{equation}
where $L_P$ is the Planck length and $L$ as in (\ref{e1}) is of the order
of the dimension under consideration.\\
The space time foam referred to above arises at the Planck scale because
the right hand side in (\ref{e2}) becomes unity, indicating perpetual
collapse and creation.\\
From this point of view, as Wheeler points out our space time is an approximation,
an average swathe at the Planck scale of several probable spaces and
topologies which form the super
space (Cf.ref.\cite{r2}). There is an immediately parallel in terms of the Compton wavelength
considerations also: As pointed out by Nottale, Abbot-wise, El Naschie, the
author and others\cite{r11,r12,r13,r14} the Quantum behaviour below a critical length is fractal and as
pointed out by the author\cite{r8}, our space time is
the thick brush stroke of thickness of the order of the Compton wavelength
of a jagged, fractal coastline like underpinning.\\
In the light of the above considerations the fluctuational creation of
particles considered by Hayakawa\cite{r15} and the author\cite{r5} have a parallel
in the non local worm hole related appearance of particles and fields at
the Planck scale\cite{r2}.\\
We will now quantify the above parallels and show the actual emergence of
the Planck scale particles from the Compton wavelength considerations.\\
We first observe that in an actual random heap of particles, the smaller
particles (in our case those having smaller Compton wavelengths and therefore
higher mass) tend to settle down together due to gravity. In a fluctuationally
created random heap of particles, there is no gravity, but as this space
time heap is not only non differentiable, but is also not required to be even
a continuum the random motion would have a similar effect: Of the
$N' = \sqrt{N}$ particles which are less dispersed, $\sqrt{N'}$ particles would
similarly fluctuationally, that is non locally be together. This fluctuationally
bound group would have a mass $\sqrt{N'}m \sim 10^{-5}gms$ or the Planck mass,
since $m$ is the mass of the pion. (Cf.ref."Ramification" for another interesting
perspective).\\
One way of looking at this is that in the above scenario, space time no
longer has the rigid features of Classical and Quantum Physics - on the average
it is a measure of dispersion of a random distribution of particles which
themselves have a stochastic underpinning. So the length scale or dispersion
would be less, the less dispersed the random collection of particles is -
this leads to the Planck scale from the Compton scale. However it must be
borne in mind that a Planck mass has a life time $\sim 10^{-42}$ seconds, and
can hardly be detected.\\
The Planck scale corresponds to the extreme classical limit of
Quantum Mechanics, as can be immediately seen from the fact that the Planck
mass $m_P \sim 10^{-5}gms$ corresponds to a Schwarzchild Black Hole of
radius $L_P \sim 10^{-33}cms$, the Planck length. At this stage the spinorial
Quantum Mechanical feature as brought out by the Kerr-Newman type Black
Hole and the Compton wavelength
(Cf.detailed discussion in refs.\cite{r3,r4}) disappears. Infact at
the Planck scale we have
\begin{equation}
\frac{Gm_P}{c^2} = \hbar /m_Pc\label{e3}
\end{equation}
In (\ref{e3}), the left side gives the Schwarzchild radius while the right
side gives the Quantum Mechanical Compton wavelength. Another way of writing
(\ref{e3}) is,
\begin{equation}
\frac{Gm^2_P}{e^2} \approx 1,\label{e4}
\end{equation}
Equation (\ref{e4}) expresses the well known fact that at this scale the entire energy is gravitational,
rather than electromagnetic, in contrast to equation (\ref{e1}) for a typical
elementary particle mass, vi.,
$$Gm^2 \approx \frac{1}{\sqrt{N}} e^2 \sim 10^{-40}e^2$$
Interestingly from the background ZPF, Planck particles can be produced at the
Planck scales given by (\ref{e3}), exactly as in the case of pions, as seen
earlier. They have been considered to be what may be called a Zero Point
Scale\cite{r17,r18,r19}. But these shortlived Planck
particles can at best describe a space time foam.\\
We will now throw further light on the fact that at the Planck scale it is
gravitation alone that manifests itself. Indeed Rosen\cite{r20} has pointed
out that one could use a Schrodinger equation with a gravitational interaction
to deduce a mini universe, namely the Planck particle. The Schrodinger
equation for a self gravitating particle has also been considered\cite{r21},
from a different point of view. We merely quote the main
results.\\
The energy of such a particle is given by
\begin{equation}
\frac{Gm^2}{L} \sim \frac{2m^5G^2}{\hbar^2}\label{e5}
\end{equation}
where
\begin{equation}
L = \frac{\hbar^2}{2m^3G}\label{e6}
\end{equation}
(\ref{e5}) and (\ref{e6}) bring out the characteristic of the Planck
particles and also the difference with elementary particles, as we will
now see.\\
We first observe that for a Planck mass, (\ref{e5}) gives, self consistently,
$$\mbox{Energy} \quad = m_P c^2,$$
while (\ref{e6}) gives,
$$L = 10^{-33}cms,$$
as required.\\
However, the situation for pions is different: They are parts of the universe
and do not constitute a mini universe. Indeed, if, as above there are $N$ pions
in the universe, then the total gravitational energy is given by, from
(\ref{e5}),
$$\frac{NGm^2}{L}$$
where now $L$ stands for the radius of the universe $\sim 10^{28}cm$.
As this equals $mc^2$, we get back as can easily be verified,
the pion mass!\\
Indeed given the pion mass, one can verify from (\ref{e6}) that $L = 10^{28}cms$
which is the radius of the universe, $R$. Remembering that $R \approx \frac{c}{H}$,
(\ref{e6}) infact gives back the supposedly mysterious and adhoc Weinberg formula,
relating the Hubble constant to the pion mass\cite{r22}.\\
This provides a justification for taking a pion as a typical particle of the
universe, and not a Planck particle, besides re-emphasizing the basic unified
picture of gravitation and electromagnetism. It must be mentioned that just
as the Planck particle constitutes a mini universe or Black Hole, so also the
$N \sim 10^{80}$ pion filled universe can itself considered to be a Black
hole\cite{r23}!\\
To proceed, let us now use the fact that our minimum space time intervals are $(l_P, \tau_P)$,
the Planck scale, instead of $(l, \tau)$ of the pion, as above.\\
With this new limit, it can be easily verified that the total mass in the volume
$\sim l^3$ is given by
\begin{equation}
\rho_P \times l^3 = M\label{e7}
\end{equation}
where $\rho_P$ is the Planck density and $M$ is the mass of the
universe.\\
Moreover the number of Planck masses in the above volume $\sim l^3$ can
easily be seen to be $\sim 10^{60}$. However, it must be remembered that
in the physical time period $\tau$, there are $10^{20}$ (that is $\frac{\tau}
{\tau_P})$ Planck life times. In other words the number of Planck particles
in the physical interval $(l, \tau)$ is $N \sim 10^{80}$, the total particle
number, as if all these were the seeds of the fixed number of $N$ particles
in the universe. This is symptomatic of the fact that instead of the elementary
particle Compton wavelength scale of the physical universe we are using the
Planck scale (cf. also considerations before equation (\ref{e3})).\\
That is from the typical physical interval $(l, \tau)$ we recover the entire
mass and also the entire number of particles in the universe, as in the Big
Bang theory. This also provides the explanation for the above puzzling
relations like (\ref{e7}).\\
That is the Big Bang theory is a characterization of the new Compton
wavelength model in the
classical limit at Planck scales, but then, in this latter case we cannot deduce from theory the
relations like the Dirac coincidences or the Weinberg formula.\\
In the spirit of\cite{r7}, one can now see the semi-classical and Quantum
Mechanical divide between Planck particles and elementary particles in the
following way. We will see that Planck particles have a life time given by
the Hawking Radiation Law of Black Hole Thermodynamics, whereas elementary
particles are characterised by Quantum Mechanical life times.\\
It is well known that\cite{r24} the life time due to the Hawking Radiation
Law is given by
\begin{equation}
t = \frac{G^2m^3}{\hbar c^4}\label{e8}
\end{equation}
which for the Planck particles gives the usual Planck time.\\
However this formulation is not valid for elementary particles. In this case,
we consider the gravitational energy $\Delta E$ of a pion as given by
an equation like (\ref{e5}) and use instead the Quantum Mechanical relation
\begin{equation}
\Delta E. \Delta t \sim \hbar\label{e9}
\end{equation}
to get
\begin{equation}
Gm^2_\pi (\hbar/m_\pi c) \Delta t \sim \hbar\label{e10}
\end{equation}
which is correct if in (\ref{e9}) $\Delta t \sim \frac{1}{H}$, the age of the universe!
(cf.also ref.\cite{r24})). In this case equation (\ref{e10}) gives the well
known and supposedly mysterious and empirical formula of Weinberg referred to earlier, viz.,
\begin{equation}
m^3_\pi \sim \frac{H \hbar^2}{Gc}\label{e11}
\end{equation}
One way of looking at this is that it is the emergence of Quantum
Mechanical effects and electromagnetism at the Compton wavelength scales
from classical gravitational considerations at the Planck scale as seen
above, which gives stability to the universe as expressed by (\ref{e9})
and (\ref{e10}).\\
All this has been justified from stochastic considerations\cite{r7}.\\
Another way of looking at all this is the following: The gravitational constant
$G$ is taken to be a universal constant in most conventional theories.
However in the above formulation it turns out that,
\begin{equation}
G = \frac{G_0}{\sqrt{N}} \propto \frac{1}{T}\label{e12}
\end{equation}
where $N$ is the number of elementary particles in the universe and $T$ is the
age of the universe. This time varying gravitational constant can be shown
to lead to consistent results including an explanation for the all important
precision of the perihelion of the Planet Mercury \cite{r6,r25}. The equation
(\ref{e12}) also shows a Machian or holistic character. In any case for a
single particle universe, $N = 1$ the $G$ above leads to the Planck
length or Planck mass, while for $N \sim 10^{80}$ the same equation leads
to the pion Compton wavelength and the usual Physics and Cosmology. Infact
if the pion Compton time scales $(l, \tau)$ tends to zero or the Planck
scale we recover the
big bang scenario and the usual space time of Classical and Quantum Physics
or the
Prigogine Cosmology\cite{r26}. In these cases we
cannot explain the large number "coincidences" and Weinberg's mysterious
formula (\ref{e11}), whereas at the elementary particle Compton scale these
features can be deduced as consequences of the theory.


\begin{thebibliography}{99}
\bibitem {r1} Misner, C.W., Thorne, K.S., and Wheeler, J.A., (1973), Gravitation,
Freeman (San Francisco).
\bibitem {r2} Wheeler, J.A., (1968) "Superspace and the Nature of Quantum
Geometrodynamics", Battelles Rencontres, Lectures, Eds., B.S. De Witt and
J.A. Wheeler, Benjamin, New York.
\bibitem {r3} Sidharth, B.G., (1998) Int.J. of Mod.Phys.A
13(15), pp2599ff.
\bibitem {r4} Sidharth, B.G., (1998) Gravitation \& Cosmology, 4 (2) (14), 158ff.
\bibitem {r5} Sidharth, B.G., (1998) International Journal of Theoretical
Physics, {\bf 37 (4)}, 1307-1312.
\bibitem {r6} Sidharth, B.G., "Effects of Varying G" to appear in Nuovo Cimento B.
\bibitem {r7} Sidharth, B.G., "Universe of Chaos and Quanta", in Chaos,
Solitons and Fractals, in press. xxx.lanl.gov.quant-ph: 9902028.
\bibitem {r8} Sidharth, B.G., "Space Time as a Random Heap", to appear in Chaos,
Solitons and Fractals.
\bibitem {r9} De Pena, L., (1983), in "Stochastic Processes Applied to Physics", Ed.,
B. Gomez., World Scientific, Singapore.
\bibitem {10} Haisch, B., Rueda, A., and Puthoff, H.E., (1994), Phys. Rev.
\underline{A49}(2), pp 678-694.
\bibitem {r11} L. Nottale, (1994), Chaos, Solitons \& Fractals, 4, 3, 361-388
and references therein.
\bibitem {r12} Abbott L.F., and Wise, M.B., (1981), AMJ Phys., \underline{49}, 37-39.
\bibitem {r13} El Naschie, M.S., (1993), Vastas Astr. \underline{37}, 249-252.
\bibitem {r14} Sidharth, B.G., (1999) "Dimensionality and Fractals", Invited
contribution to Special Issue of Chaos, Solitons and Fractals on 'Fractal
Geometry in Quantum Physics'.
\bibitem {r15} Hayakawa, S., (1965), Suppl of PTP Commemmorative Issue, 532-541.
\bibitem {r16} Sidharth, B.G., "Quantum Mechanical Black Holes: Issues and Ramifications",
to appear in Proceedings of 'Frontiers of Fundamental Physics'.
\bibitem {r17} Winterberg, F., Intl.J.Th.Phys. \underline{33}(6), (1994).
\bibitem {r18} Padmanabhan, T., IUCAA Preprint 1/98, (1998).
\bibitem {r19} Isham, C.J., Kubyshin, Y., and Renteln, P., Class.Quantum Grav.
(1990), 7, 1053-1074.
\bibitem {r20} Rosen, N., (1993) International Journal of Theoretical Physics,
32 (8), 1435-1440.
\bibitem {r21} Sidharth, B.G., and Popova, A.D., (1996), Differential Equations
and Dynamical Systems, 4 (3/4), 431-440.
\bibitem {r22} Weinberg, S., (1972), Gravitation and Cosmology, Wiley, New York.
\bibitem {r23} Sidharth, B.G., (1999) "Fluctuational Cosmology" in Quantum Mechanics
and General Relativity" in Proceeding of the
Eighth Marcell Grossmann Meeting on General Relativity, Ed., T. Piran, World
Scientific, Singapore, pp.476ff.
\bibitem {r24} C. Sivaram, (1982), Astrophysics and Space Science, \underline{88}.
\bibitem {r25} Sidharth, B.G., "Further Effects of Varying G", xxx.lanl.gov Phys 0001062.
\bibitem {r26} Sidharth, B.G., (1999) Ast \& Geophysics (Journal of Royal Astronomical
Society), 40, p.2.8.
\end{thebibliography}
\end{document}